\documentclass[aps,prd,twocolumn,superscriptaddress,nofootinbib,10pt]{revtex4-1}
\usepackage[paperwidth=21cm,paperheight=29.7cm,top=2.54cm,bottom=2.54cm,left=2cm,right=2cm]{geometry}
\usepackage[colorlinks,linkcolor=blue,anchorcolor=blue,citecolor=blue,urlcolor=blue]{hyperref}
\usepackage{graphicx,subfigure}
\usepackage[figuresright]{rotating}
\usepackage{amsmath,amsfonts,amssymb,bm}
\usepackage{array,enumitem,multirow}
\usepackage{acronym}
\usepackage{makecell}
\newcommand{\be}{\begin{equation}}
\newcommand{\ee}{\end{equation}}
\newcommand{\bea}{\begin{eqnarray}}
\newcommand{\eea}{\end{eqnarray}}

\newacro{GR}{general relativity}
\newacro{GW}{gravitational wave}
\newacro{MG}{modified gravity theory}
\newacro{BH}{Black hole}
\newacro{PN}{post-Newtonion}
\newacro{ppE}{parameterized post-Einsteinian}
\newacro{GCB}{galactic ultra-compact binary}
\newacro{SBHB}{stellar-mass black hole binary}
\newacro{MBHB}{massive black hole binary}
\newacro{BHB}{black hole binary}
\newacro{IMBHB}{intermediate-mass black hole binary}
\newacro{EMRI}{extreme mass ratio inspiral}
\newacro{IMRI}{intermediate mass ratio inspiral}
\newacro{SGWB}{stochastic gravitational wave background}
\newacro{MECO}{minimal energy circular orbit}
\newacro{FAR}{false alarm rate}
\newacro{CE}{Cosmic Explorer}
\newacro{ET}{Einstein Telescope}
\newacro{LISA}{Laser Interferometer Space Antenna}
\newacro{EdGB}{Einstein-dilaton Gauss-Bonnet}
\newacro{dCS}{dynamic Chern-Simons}
\newacro{SNR}{signal-to-noise ratio}
\newacro{FIM}{Fisher Information Matrix}
\newacro{ISCO}{innermost stable circular orbit}
\newacro{NSBH}{neutron star-black hole binary}
\newacro{MCMC}{Markov Chain Monte Carlo}
\newacro{QNM}{quasi-nomral mode}

\begin{document}

\title{ Quasinormal modes of spontaneous scalarized Kerr black holes}

\author{Wei Xiong}
 \email{202210187053@mail.scut.edu.cn}%
\author{Peng-Cheng Li}%
 \email{pchli2021@scut.edu.cn, corresponding author}
\affiliation{%
School of Physics and Optoelectronics, South China University of Technology, Guangzhou 510641, People’s Republic of China.
}%

\date{\today}

\begin{abstract}	
	Recent studies have shown that rotating black holes can undergo spontaneous scalarization, leading to deviations from general relativity in the strong-field regime. We present the first nonperturbative calculation of the quasinormal modes (QNMs) of scalarized Kerr black holes in Einstein-scalar-Gauss-Bonnet gravity, without assuming small spin or weak coupling. Our results reveal a universal splitting of the fundamental $l=m=2$ mode into axial-led, polar-led, and scalar-led branches, breaking the isospectrality characteristic of Kerr black holes. This splitting offers distinct signatures in the ringdown phase of gravitational wave signals and provides a new avenue to test gravity in the strong-field regime. Our findings open the possibility of probing beyond-GR physics using precision measurements of black hole ringdowns in upcoming gravitational wave observations.

\end{abstract}

\maketitle



{\textit{Introduction}} ---\label{section1}
The advent of gravitational wave (GW) astronomy has opened a new window into the strong-field regime of gravity, enabling direct tests of general relativity (GR) in previously inaccessible regimes~\cite{LIGOScientific:2018mvr,Berti:2015itd}. Among GR’s most striking predictions, black holes (BHs) serve as ideal laboratories due to their simplicity and universality~\cite{Barack:2018yly}. In GR, the no-hair theorem asserts that stationary BHs are fully characterized by mass, spin, and electric charge~\cite{Herdeiro:2015waa}. Precise measurements of GW signals from binary mergers now allow tests of this paradigm, including potential deviations arising from additional scalar or other matter fields~\cite{Isi:2019aib,Barausse:2020rsu,Brito:2017wnc,Hannuksela:2018izj}, offering a powerful probe of physics beyond Einstein’s theory.


Spontaneous scalarization is a nonlinear mechanism through which BHs can acquire scalar hair in the nonlinear regime via coupling between a scalar field and a source term, while remaining consistent with weak-field tests~\cite{Antoniou:2017acq,Doneva:2017bvd,Antoniou:2017hxj,Doneva:2018rou,Collodel:2019kkx,Herdeiro:2020wei,Silva:2017uqg,Berti:2020kgk,Herdeiro:2018wub,Silva:2017uqg,Cunha:2019dwb}. Among various realizations, Einstein-scalar-Gauss-Bonnet (EsGB) gravity has emerged as a particularly compelling framework, where the scalar field couples to the GB term---a high-energy-motivated curvature invariant free from Ostrogradsky instabilities~\cite{Zwiebach:1985uq,Ostrogradsky:1850fid}. In this theory, scalarization is triggered when Kerr BHs become unstable against scalar perturbations beyond a critical coupling, leading dynamically to the formation of scalarized rotating BHs~\cite{Dima:2020yac,Doneva:2020nbb,East:2021bqk}. These solutions closely mimic Kerr geometry but can exhibit observable deviations, especially in their GW emission. Scalarized BHs in EsGB theory have been studied in contexts ranging from shadows~\cite{Cunha:2019dwb} and dynamical formation~\cite{East:2021bqk} to stability and perturbative properties~\cite{Blazquez-Salcedo:2020rhf,Silva:2018qhn,Macedo:2020tbm,Antoniou:2024gdf}. For a recent review, see~\cite{Doneva:2022ewd}.

A natural avenue to test EsGB gravity is through GW observations of coalescing compact binaries, where scalarization can leave distinct imprints. Current constraints focus primarily on the inspiral stage, where the absence of scalar dipole radiation imposes bounds on the coupling parameters~\cite{Wong:2022wni}. However, the merger and ringdown phases remain largely unexplored due to the lack of accurate waveform models in this regime.

The ringdown signal is governed by a superposition of damped sinusoids—quasinormal modes (QNMs)---whose complex frequencies encode the properties of the remnant black hole~\cite{Kokkotas:1999bd,Berti:2009kk,Konoplya:2011qq}. To probe EsGB deviations during ringdown, the QNM spectrum of rotating scalarized BH must first be computed. Previous studies have addressed the static case~\cite{Blazquez-Salcedo:2020rhf,Blazquez-Salcedo:2020caw}, but extending to rotating backgrounds remains challenging: the scalarized Kerr geometry lacks Petrov type D symmetry, invalidating the standard Teukolsky formalism~\cite{Teukolsky:1973ha}.
Recent advances have explored alternative methods to compute QNMs in modified gravity, including generalized Teukolsky approaches~\cite{Li:2022pcy,Hussain:2022ins}, full numerical relativity simulations~\cite{Okounkova:2019zjf,AresteSalo:2022hua}, and small-spin approximations~\cite{Wagle:2021tam,Pierini:2022eim}. Notably, a spectral decomposition framework has emerged as a powerful tool to solve coupled perturbation equations for rotating black holes, yielding highly accurate QNMs in a fully nonperturbative manner~\cite{Chung:2023zdq,Chung:2023wkd,Blazquez-Salcedo:2023hwg,Chung:2024ira,Blazquez-Salcedo:2024oek}.

In this work, we present the first fully nonperturbative calculation of the quasinormal mode (QNM) spectrum for rotating scalarized black holes in EsGB gravity. Utilizing spectral decomposition, we numerically construct both the scalarized background and its linear perturbations, and recast the QNM problem as a generalized eigenvalue system. For the first time, we provide a compact and universal formulation of the dissipative boundary conditions at the event horizon and at spatial infinity. Focusing on the $l = m = 2$ fundamental mode---the dominant ringdown excitation in nearly equal-mass binary mergers confirmed by numerical relativity~\cite{Buonanno:2006ui,Berti:2007fi}---we find that it generically splits into three distinct families: axial-led, polar-led, and scalar-led. This spectral splitting breaks the isospectrality of Kerr BHs in GR and introduces potential observational signatures of BH scalarization. These findings provide a concrete, model-specific prediction of spontaneous scalarization that could be tested in future GW observations. Our work thus opens new possibilities for probing beyond-GR physics through precision BH spectroscopy.

{\textit{The EsGB model}} ---
The EsGB theory is described by the action
\begin{equation}
	S = \frac{1}{16\pi} \int d^{4} x \sqrt{-g} [R-2\nabla_{\mu} \nabla^{\mu} \phi  + \lambda^{2} f(\phi) \mathcal{R}_{\textrm{GB}}],
	\label{eq:action}
\end{equation}
where the  GB term $\mathcal{R}_{\textrm{GB}} \equiv R_{\mu\nu\rho\sigma} R^{\mu\nu\rho\sigma} -4 R_{\mu\nu} R^{\mu\nu}+R^{2}$ is coupled to a function $f(\phi)$ of the scalar field $\phi$ via the coupling constant $\lambda$. The equations of motion for the backgroud and the perturbation field can be expressed as
\begin{eqnarray}
	0 &=& \mathcal{E}_{\mu\nu}^{(0)} + \epsilon \mathcal{E}_{\mu\nu}^{(1)} + \mathcal{O}(\epsilon)^{2}, \nonumber \\
	0 &=& \mathcal{S}^{(0)} +  \epsilon \mathcal{S}^{(1)} + \mathcal{O}(\epsilon)^{2},
	\label{eq:equations}
\end{eqnarray}
where $\mathcal{E}_{\mu\nu}$ is the tensor of modified Einstein equation and $\mathcal{S}$ represents the scalar field equation. The superscripts $(0)$ and $(1)$ mark the equations for the background fields and the first-order perturbed fields,  respectively. 
The coupling function $f(\phi)$ for the spontaneous scalarization model typically satisfies the conditions $f'(0)= 0$ and $f''(0) \neq 0$. We adopt the general form of $f(\phi)$ as 
\begin{equation}
	f(\phi) =\frac{1}{12} (1-e^{-6\phi^{2}}),  
	\label{eq:CouplingFunction}
\end{equation}
by following \cite{Doneva:2017bvd,Cunha:2019dwb,Wong:2022wni}. The existence domain \cite{Cunha:2019dwb,Wong:2022wni} indicates that BHs with moderate spin ($\chi \approx 0.6 $, $\chi$ is the dimensionless spin of BHs) or lower must undergo scalarization if their characteristic scale is comparable to $\lambda$. In contrast, rapidly rotating BHs remain effectively indistinguishable from Kerr BHs predicted by GR. The coupling function with the form of (\ref{eq:CouplingFunction}) is known as the curvature-induced scalarization. Another model called spin-induced scalarization predicts scalarization exclusively for rapidly rotating BHs, achieved by simply multiplying the coupling function in (\ref{eq:CouplingFunction}) by $-1$ \cite{Herdeiro:2020wei,Berti:2020kgk}. 

{\textit{The numerical method}} 
\begin{figure}[htbp]
	\centering
	\includegraphics[width = 0.450\textwidth]{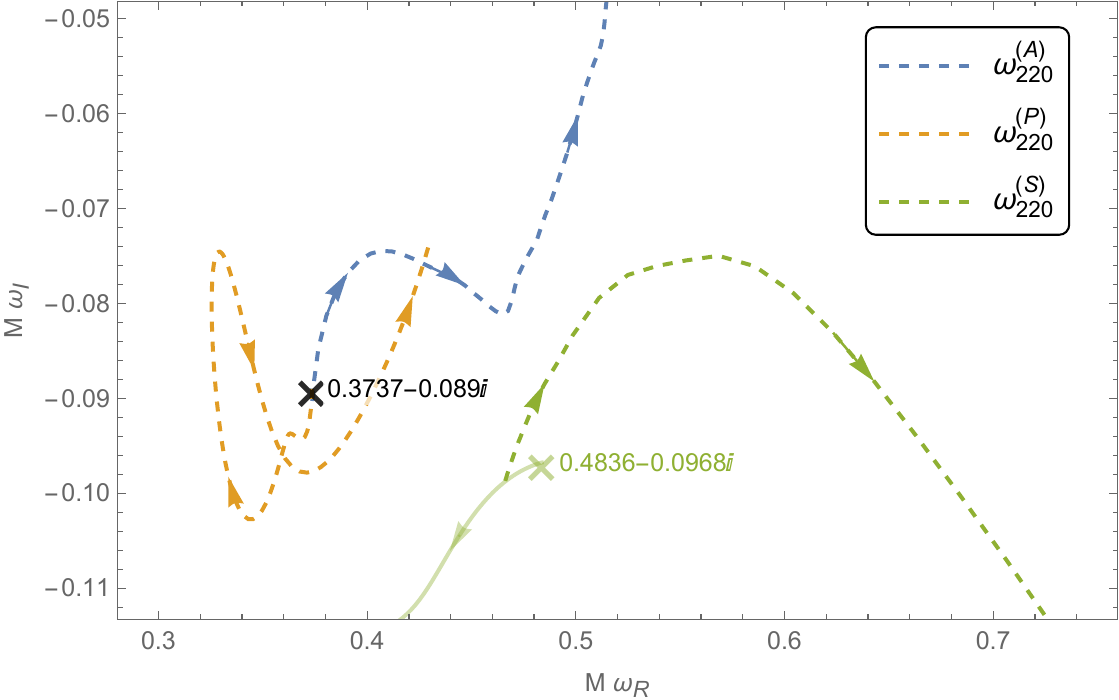}
	\caption{The migration of the $l=m=2$ fundamental mode (dashed lines) for scalarized static BHs on the complex plane with increasing $\lambda/M$ described by the direction of arrows. The black (green) crosses denotes the mode for the gravitational perturbation (the scalar perturbation) at $\lambda=0$. As $\lambda/M$ increases, the gravitational mode of the vacuum BH remains unchanged, and the scalar mode varies as depicted by the green solid line.
	Scalarized BHs branch from the threshold of nonvanishing $\lambda/M \approx 1.704$, and hence the scalar mode for the scalarized branch (the green dashed line) emerges from the green solid line at the point $\omega \approx 0.4661- 0.0988 i$, instead the green cross (about $0.4836-0.0968 i$). 
	The other dashed lines describe gravitational modes (axial-led modes (blue) and polar-led modes (yellow)) branching from the black cross, where $\lambda/M$ ranges from $1.704$ to $6$. These results exhibit the resemblance of the perturbative response between the vacuum BH and the scalarized BH at the threshold, analogous to their thermodynamic behavior \cite{Doneva:2017bvd}.
	}
	\label{fig0}
\end{figure}
The metric ansatz for the background spontaneous scalarized Kerr BHs can be written as 
\cite{Cunha:2019dwb,Herdeiro:2020wei}
\begin{eqnarray}
	ds^{2} &=&-e^{2 F_{0}} N dt^{2} + e^{2F_{1}} \left( dr^{2}/N +r^{2} d\theta^{2} \right) \nonumber \\
	&& + e^{2F_{2}} r^{2} \sin^{2}\theta (d\varphi-W/r^{2}dt)^{2},
	\label{eq:ansatz}
\end{eqnarray}
with $N\equiv 1-r_{H}/r$ and the event horizon radius $r_{H}$. The components ($F_{0},F_{1},F_{2},W$) together with the backgroud scalar field $\phi^{(0)}$, depend on $r$ and $\theta$ only, are decomposed through a spectral method developed by \cite{Fernandes:2022gde}. The large system of algebraic equations for the coefficients, generated from substituting the spectral decomposition into the backgroud equations $\mathcal{E}_{\mu\nu}^{(0)} = \mathcal{S}^{(0)} = 0$, is solved by the Newton-Raphson method with an appropriate seed. We recommend interested readers to refer to \cite{Fernandes:2022gde} for further details. The generic perturbations of the scalarized BHs  can be separated into axial-led(A) and polar-led (B) part \cite{Blazquez-Salcedo:2023hwg}
\begin{equation}
	g_{\mu\nu} = g^{(0)}_{\mu\nu} + e^{-i\omega t+ i m \varphi} \epsilon(h_{\mu\nu}^{\textrm{(A)}} + h_{\mu\nu}^{\textrm{(P)}}),
	\label{eq:perturbedmetric}
\end{equation}
where the separation of variables $e^{-i\omega t+ i m \varphi}$ with frequency $\omega = \omega_{R}+\omega_{I} i$ and azimuth number $m$ is generated from the symmetries of the backgroud spacetime. The axial and polar perturbations are, respectively, given by
 
\begin{equation}
	h_{\mu\nu}^{\textrm{(A)}} =
	\left(
	\begin{array}{cccc}
		0& 0 & -im h_{0}/\sin \theta & \sin \theta \partial_{\theta} h_{0} \\
		0& 0 & -im h_{1}/\sin \theta &  \sin \theta \partial_{\theta} h_{1} \\
		-im h_{0}/\sin \theta & \sin \theta \partial_{\theta} h_{0} & 0 & 0 \\
		-im h_{1}/\sin \theta &  \sin \theta \partial_{\theta} h_{1} & 0 & 0 
	\end{array}
	\right),
	\label{eq:pmetricA}
\end{equation}
and 
\begin{equation}
	h_{\mu\nu}^{\textrm{(P)}} =
	\left(
	\begin{array}{cccc}
		H_{0}e^{-2F_{1}}N & H_{1} & 0 & 0 \\
		H_{1}& H_{2}e^{2F_{1}}/N & 0 & 0 \\
		0 & 0 & K e^{2F_{1}} r^{2} & 0 \\
		0 & 0 & 0 & K e^{2F_{2}} r^{2} \sin^{2} \theta
	\end{array}
	\right).
	\label{eq:pmetricP}
\end{equation}
The components ($h_{0},h_{1},H_{0},H_{1},H_{2},K$) for the perturbed metric are straightforwardly extended from the Regge-Wheeler gauge and the Zerilli gauge. However, the axial and polar functions do not decouple from each other in a general axisymmetric spaceetime.

\begin{figure*}[t!]
	\begin{center}
		\mbox{ 
			\includegraphics[width = 0.488\textwidth]{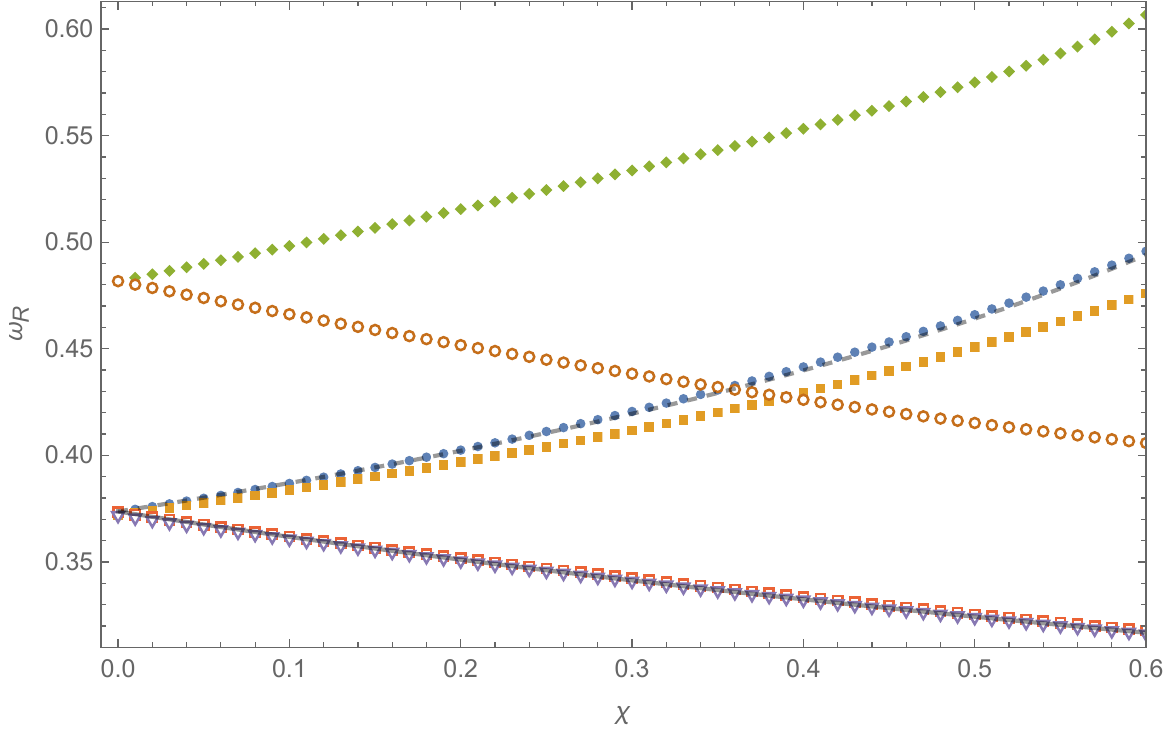}
			\includegraphics[width = 0.50\textwidth]{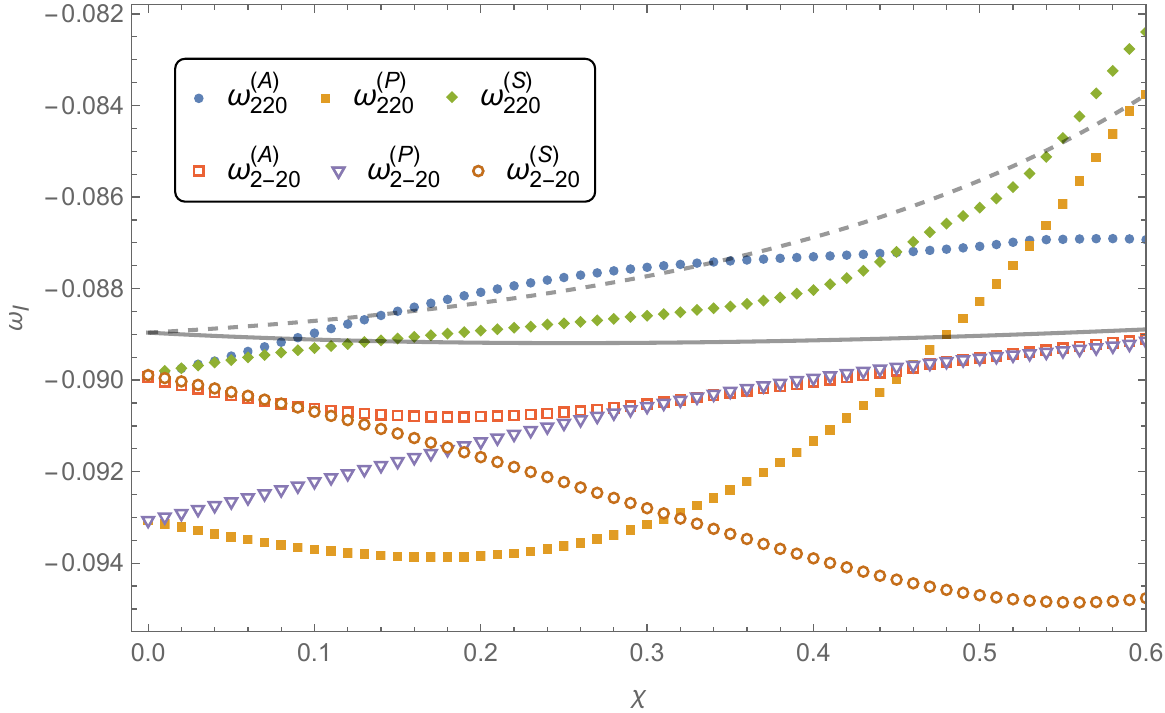}
		}
		\vspace*{-0.5cm}
	\end{center}
	\caption{
		The spectrum of the fundamental $l=|m|=2$ QNMs for scalarized Kerr BHs are depicted while varying BH spin $\chi$ and fixing the coupling constant $\lambda/M = 1.75$. 
		The left panel shows the real part of the QNMs with varying spin, while the right panel displays the imaginary part. Solid and hollow markers represent the prograde and retrograde mode of the scalarized BHs, respectively. The different colors of markers corresponding to different modes are listed in the legend within the right panel. 
		For comparison, the grey dashed line show the prograde modes of Kerr BHs with $m=2$, while the retrograde modes with $m=-2$ are depicted by the grey solid line.  From the right panel, the prograde families always dominate the spectrum as the mode with smallest imaginary part, analogous to the spectrum of Kerr BHs (the grey dashed line is always larger than the grey solid line). However, the polar prograde modes (the yellow solid markers) remarkably exhibit a smaller imaginary part than the polar retrograde mode (the purple hollow markers).
	}
	\label{fig1}
\end{figure*} 
We substitute the perturbed metric (\ref{eq:pmetricA}) and (\ref{eq:pmetricP}) and the perturbed scalar field $\phi^{(1)}$ into the first-order perturbation equations $\mathcal{E}_{\mu\nu}^{(1)} = \mathcal{S}^{(1)} = 0$, resulting in a set of seven coupled linear partial differential equations for seven perturbation components. The corresponding boundary conditions for the perturbation components can be written in a compact form
\begin{equation}
	f(r,\theta) \sim
	\left\{
	\begin{array}{cll}
		& (r-r_{H})^{-\frac{i(\omega-m\Omega_{H})}{2\kappa_{H}}}, & r \rightarrow r_{H}, \\
		& & \\
		& e^{i\omega r} r^{2 i M\omega } , & r \rightarrow +\infty,
	\end{array}
	\right.
	\label{eq:boundary}
\end{equation}
where $f(r,\theta)$ represents an arbitrary component of ($h_{0},h_{1},H_{0},H_{1},H_{2},K,\phi^{(1)}$), $\Omega_{H} \equiv \frac{W}{r^{2}}|_{r=r_{H}}$ is the angular velocity at the event horizon and $\kappa_{H}\equiv \frac{1}{2r_{H}} e^{F_{0}-F_{1}} |_{r=r_{H}}$ is the surface gravity. The boundary condition at the event horizon can be derived by requiring the regularity of perturbations in that region \cite{Xiong:2024}. Moreover, these conditions have been shown to be universal for all massless perturbations on the background metric given in Eq.~(\ref{eq:ansatz}) \cite{Xiong:2024}, and are consistent with those established for vacuum Kerr black holes and scalarized BHs \cite{Blazquez-Salcedo:2020rhf, Pierini:2021jxd,Chung:2024vaf}. 

To discretize the perturbation equations, we decompose the perturbation functions as 
\begin{equation}
	f(x,\theta) = \sum^{N_{x}-1}_{i=0} \sum^{N_{\theta}-1}_{j=0} c_{ij} \ C^{(x)}_{i}(x) \ C^{(\theta)}_{j}(\theta).
	\label{eq:decomposition}
\end{equation}
The cardinal function for $x$ is constructed by the linear combinations of Chebyshev polynomials
\begin{equation}
	C^{(x)}_{i}(x) =  \sum_{n=0}^{N_{x}-1} \frac{T_{n}(x_{i})T_{n}(x)}{P^{(x)}_{n}},
	\label{eq:cardinalx}
\end{equation}
where $P^{(x)}_{n}  = N_{x}$ for $n=0$ and $P^{(x)}_{n}  = N_{x}/2$ for $n \neq 0$.  $x \equiv 1- 2 r_{H}/r$ is defined by the compactification of the radial coordinate. The cardinal function for $\theta$ is
\begin{equation}
	C^{(\theta)}_{j}(\theta) = \sum_{n=0}^{N_{\theta}-1} \frac{\cos(n \ \theta_{j}) \cos(n \ \theta)}{P^{(\theta)}_{n}},
	\label{eq:cardinaltheta}
\end{equation}
for even azimuth number $m$, with  $P^{(\theta)}_{n}  = N_{\theta}$ for $n=0$ and $P^{(\theta)}_{n}  = N_{\theta}/2$ for $n \neq 0$. 
The collocation points appearing in the above cardinal functions are defined as 
\begin{equation}
	x_{i} = \cos \left( \frac{(2 i +1) \pi}{2 N_{x}} \right), \ \ \theta_{j} = \frac{(2j+1)\pi}{2N_{\theta}},
\end{equation}
with $i=0, \ldots , N_{x}-1$ and $j=0, \ldots,  N_{\theta}-1$.
The cardinal functions (\ref{eq:cardinaltheta}) for the angular part differ from the Legendre polynomials, though the latter is typically the natural basis for angular components in separable perturbations of spherically symmetric BHs \cite{Blazquez-Salcedo:2023hwg}.  However, the two choices are equivalent, as their respective basis functions can be exactly transformed into one another. 

The expansion coefficients of the spectral decomposition (\ref{eq:decomposition}) are obtained by  $c_{ij} = f(x_{i},\theta_{j})$. We utilize the co-lexicographic ordering to flatten the two-dimensional array $f(x_{i},\theta_{j})$ into a vector  $\vec{f}=\{f_{I}, I=1,2,...,N_x\times N_\theta\}$, and hence the differential operator acting on the perturbation function can be expressed as a Kronecker product \cite{Dias:2015nua}
\begin{equation}
	\frac{d^{\alpha}}{dx^{\alpha}} \frac{d^{\beta}}{d\theta^{\beta}} f(x,\theta) \rightarrow \sum_{I=1}^{N_{x} \times N_{\theta}}  f_{I} \left[[D^{(\alpha)}]^{(x)} \otimes [D^{(\beta)}]^{(\theta)}\right]_{IJ},
	\label{eq:derivativeMatrix}
\end{equation}
with the $\alpha$-derivative matrix $[D^{(\alpha)}]^{(x)}_{ij} \equiv d^{\alpha}C^{(x)}_{i}(x_{j})/dx^{\alpha}$ and $[D^{(\alpha)}]^{(\theta)}_{ij} \equiv d^{\alpha}C^{(\theta)}_{i}(\theta_{j})/d\theta^{\alpha}$ with respect to $x$ and $\theta$, respectively. 

All the steps discretize the perturbation equations $\mathcal{E}_{\mu\nu}^{(1)} = \mathcal{S}^{(1)} = 0$ into a matrix equation $(\mathcal{M}_{0}+\omega \mathcal{M}_{1}+\omega^{2} \mathcal{M}_{2})\vec{f}=0$, where $\mathcal{M}_{0},\mathcal{M}_{1},\mathcal{M}_{2}$ represent rectangular matrices of order ($7 \times N_{x} \times N_{\theta}$). The QNM frequency $\omega$ can be solved as a generalized eigenvalue problem for this matrix equation. 
\begin{figure*}[t!]
	\begin{center}
		\mbox{ 
			\includegraphics[width = 0.3\textwidth]{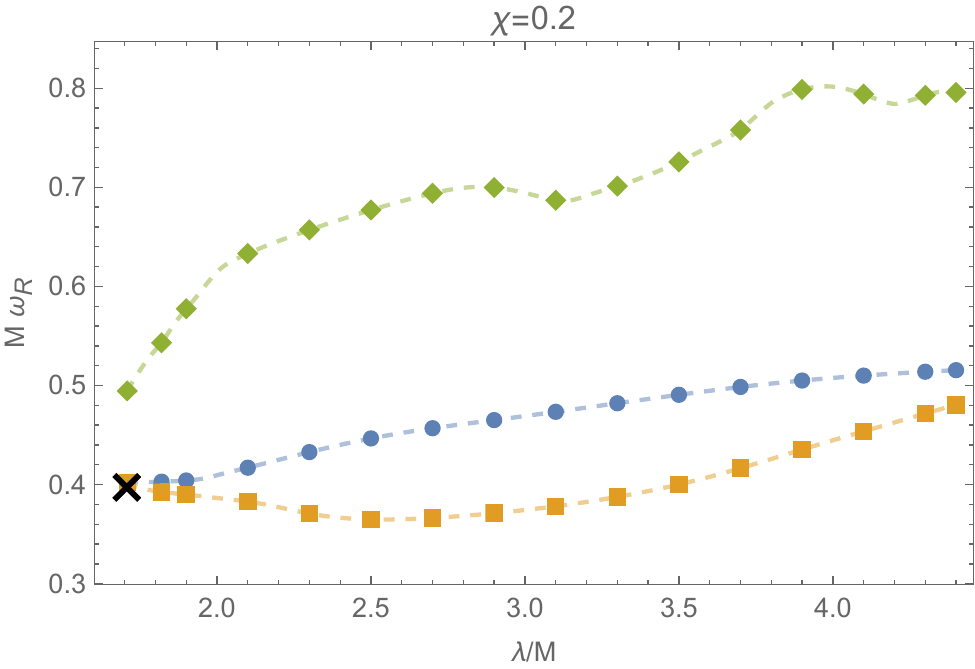}
			\includegraphics[width = 0.3\textwidth]{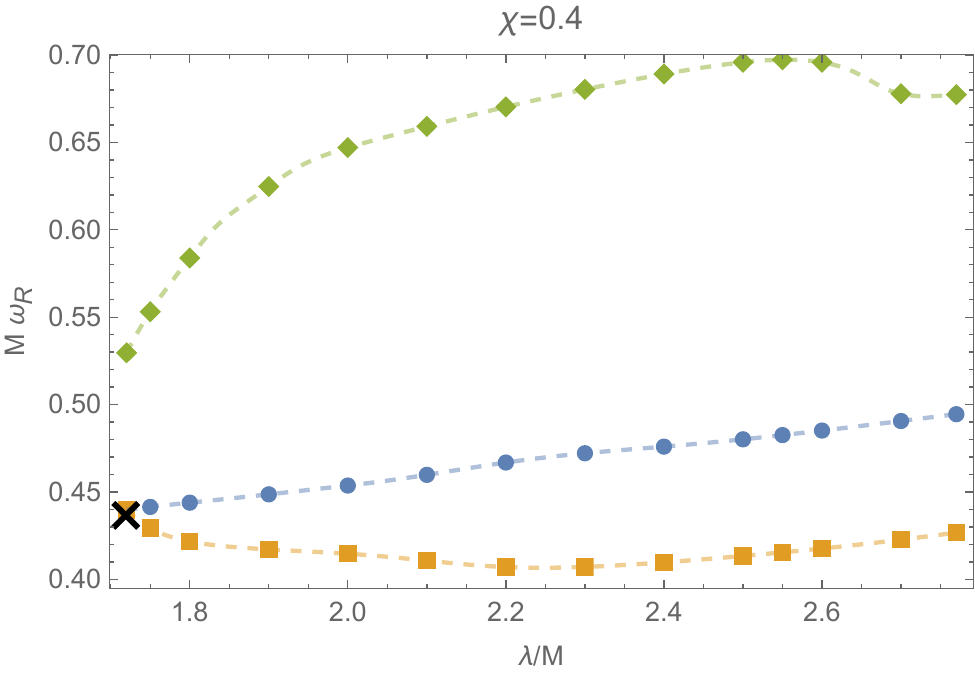}
			\includegraphics[width = 0.303\textwidth]{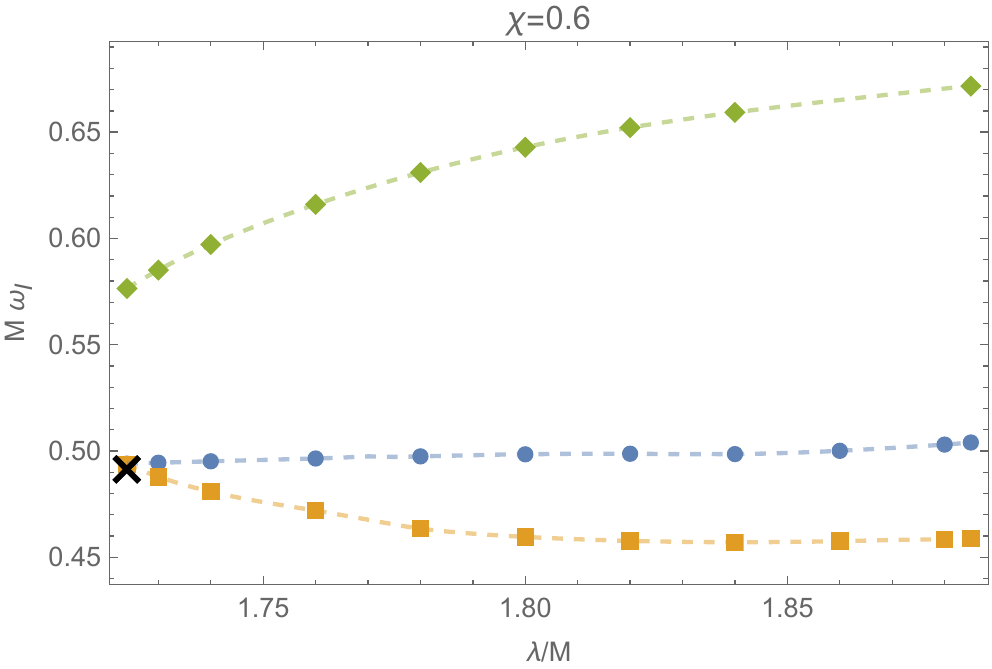}
		}
		\mbox{ 
			\includegraphics[width = 0.31\textwidth]{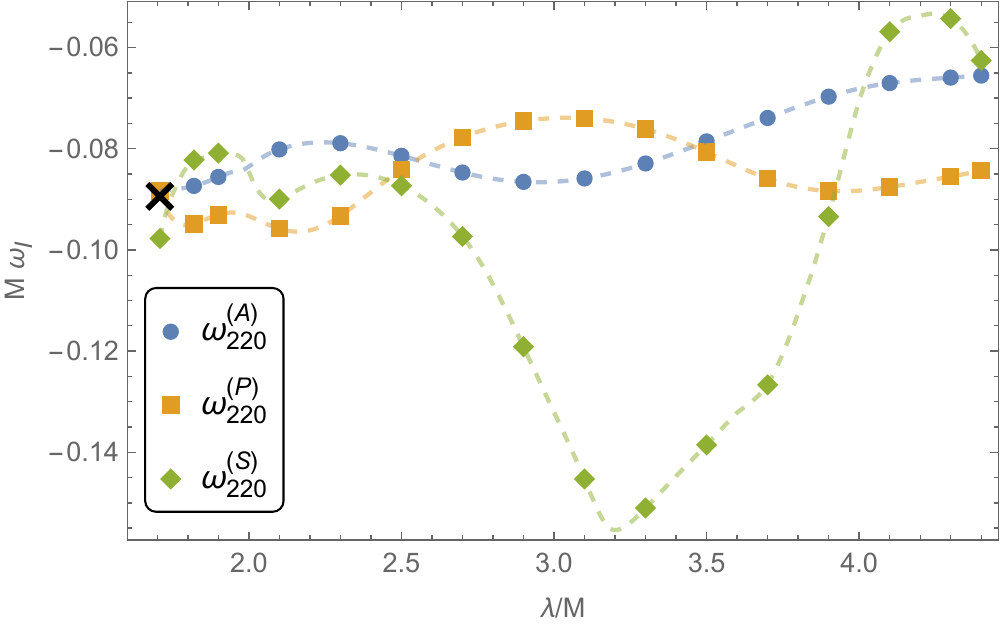}
			\includegraphics[width = 0.31\textwidth]{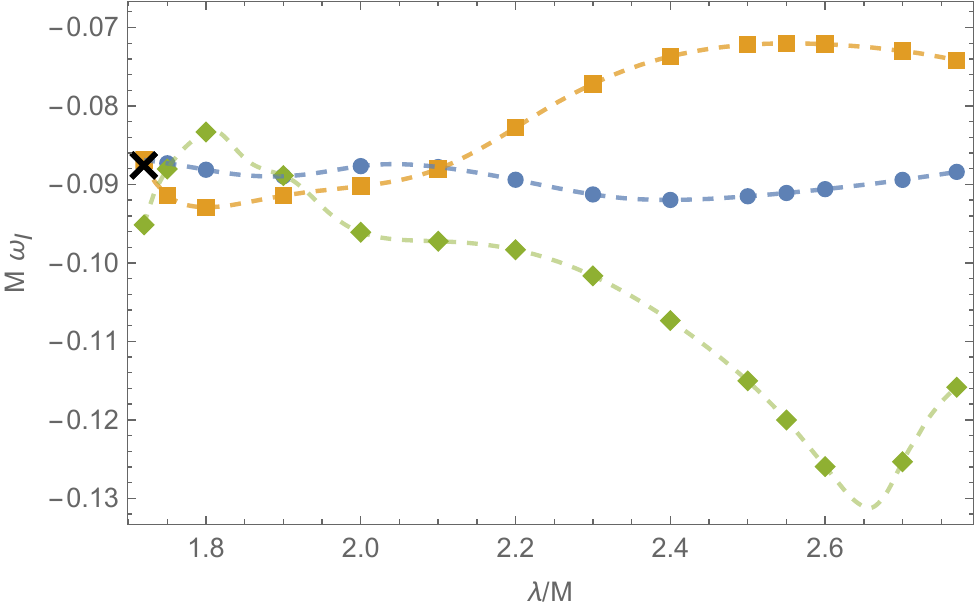}
			\includegraphics[width = 0.318\textwidth]{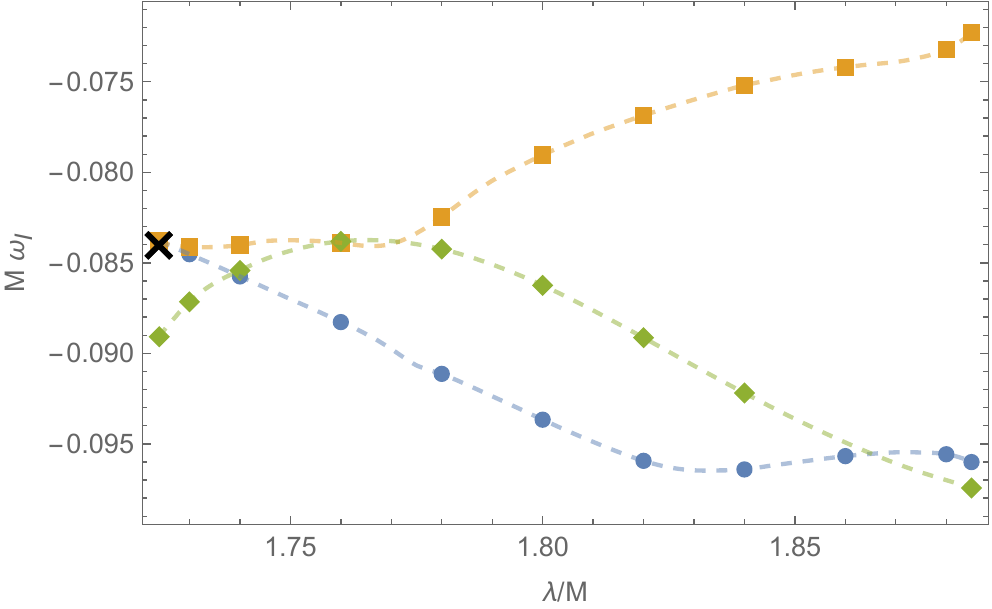}
		}
		\vspace*{-0.5cm}
	\end{center}
	\caption{
		The $l=m=2$ prograde modes with varying dimensionless coupling constant $\lambda/M$ for rotating scalarized BHs with different spin. The different markers denote the axial modes (blue circles), the polar modes (yellow squares) and the green modes (green diamonds), respectively, as listed in the small frame within the left bottom and held to the remaining panels. The black crosses represent the $l=m=2$ gravitational modes of Kerr BHs with following spin. The three values ($\chi=0.2,0.4,0.6$) of the fixed BH spin are shown at the top of the three column. For each column, the first row displays the real part of QNMs, while the second row shows the imaginary parts. These plots illustrate the emergence of the polar and axial modes for scalarized BHs from the gravitational modes of Kerr BHs at the critical $\lambda/M$. As $\lambda/M$ increases, the polar and axial modes split from each other into two distinct line. 
		}
	\label{fig2}
\end{figure*}

{\textit{Results}} ---
The dynamical evolution of binary BH mergers within the framework of GR shows that the most easily excited mode is the $l=|m|=2$ fundamental mode, that is the mode with largest imaginary part \cite{Buonanno:2006ui,Berti:2007fi}. In the following, we concentrate on the $l=|m|=2$ fundamental mode.

In Fig.\ref{fig0}, we show how the $l=|m|=2$ fundamental mode of the scalarized static BH branches from the one of the vacuum BHs and splits into three distinguishable families: axial, polar, and scalar modes, as the scaled coupling constant $\lambda /M$ (where $M$ is the mass) increases. Particularly, the scalar mode of scalarized BHs should branch from that of the vacuum BH at $\omega \approx  0.4661-0.0987i$ with the threshold $\lambda/M \approx 1.704$ instead of $\lambda=0$, as illustrated by the intersection of the green dashed and green solid lines in Fig.\ref{fig0}. The scalar modes of vacuum BHs are calculated using the accurate continued fraction method and validated by the spectral method. The described behavior indicates that the QNMs of scalarized BHs can asymptotically approach those of vacuum BHs while $\lambda/M$ approach the threshold, analogous to the thermodynamic behavior \cite{Doneva:2017bvd}. We reproduce the axial modes of the scalarized static BH reported in \cite{Blazquez-Salcedo:2020rhf}. Nevertheless, in Fig.\ref{fig0}, the polar and scalar modes exhibit discrepancies compared to \cite{Blazquez-Salcedo:2020caw}. We comment that the frequency value $\omega \approx  0.481-0.0894i$ of the scalar mode   of the scalarized BHs at the threshold point reported in \cite{Blazquez-Salcedo:2020caw} dose not match the value of the vacuum BHs.

Fig.\ref{fig1} illustrates the variation of the real part (left panel) and imaginary part (right panel) for the fundamental $l=|m|=2$ modes as the spin increases for scalarized BHs with $\lambda=1.75$, including both prograde modes (defined by $m \omega_{R}>0$ and denoted by solid markers) and retrograde modes (defined by $m \omega_{R}<0$ and denoted by hollow markers). 
For scalarized Kerr BHs, the prograde modes (or the retrograde modes) can further be categorized by three families (the axial, polar and scalar-led modes) according to their main perturbed components (\ref{eq:perturbedmetric}) \cite{Blazquez-Salcedo:2024oek},  which their names inherit from those of the static limit \cite{Blazquez-Salcedo:2020caw,Blazquez-Salcedo:2020rhf}. In summary, the fundamental modes for given ($l,|m|$) are categorized into six families, with the corresponding markers listed in the legend on the right panel of Fig.\ref{fig1}. we also depict the prograde modes (the grey dashed line) of Kerr BHs with $|l|=m=2$ and the retrograde modes (the grey solid line) with $|l|=m=-2$ in each panel for comparison. 
It is clear that the isospectrality, which holds for the Kerr axial-led and polar-led modes, is broken once the BHs are scalarized. 
We find an interesting behavior for the polar-led modes: for certain spin, the prograde mode (the yellow squares) can exhibit a smaller imaginary part compared to the retrograde mode (purple inverted triangles). In contrast, the behavior of the two branches for the axial-led modes resembles that of Kerr BHs, where the prograde modes always dominate. 

We demonstrate the QNMs of rotating scalarized BHs for varying coupling constant $\lambda/M$ and fixing the dimensionless spin $\chi$ in Fig.\ref{fig2}. 
Three fixed values ($\chi=0.2,0.4,0.6$) of the spin are shown at the top of each column for the panels. For each column, the first row presents the real part of the QNMs, while the second row demonstrates the imaginary part. In each panel, the blue circles, yellow squares, and green diamonds represent the axial-led, polar-led, and scalar-led modes, respectively. We depict the $|l|=m=2$ gravitational modes of Kerr black holes by the black cross in each panels, highlighting the splitting between axial-led and polar-led modes for the scalarized BHs.
At low spin $\chi=0.2$, the behavior of each family is similar to that in the static BH limit  as shown by Fig.\ref{fig0}. The polar-led mode dominates only at intermediate $\lambda/M$, while the axial-led mode dominates otherwise. For larger spin $\chi=0.4$, the polar-led mode becomes dominant over larger values of the coupling constant, and the polar-led mode can even maintain dominance for $\chi=0.6$. 
The scalar-led mode becomes dominant only at either small or large values of $\lambda/M$, with the range of dominance being relatively narrow.

{ \textit{Discussion}} ---
We have presented the first nonperturbative computation of the QNM spectrum of rotating scalarized BHs in EsGB gravity. Our approach captures both the nonlinear scalarized background and its full linear perturbations using spectral methods, allowing accurate determination of QNMs beyond previous approximations. A key result is the universal splitting of the fundamental $l=m=2$ mode into three families---axial-led, polar-led, and scalar-led---breaking the isospectrality characteristic of Kerr BHs in GR. This effect, absent in GR, introduces distinct signatures that may be detectable in the ringdown phase of GWs from BH mergers.

Our findings provide concrete theoretical predictions for the observable consequences of BH scalarization. As GW detectors continue to improve in sensitivity, precision spectroscopy of the ringdown phase may offer a novel avenue to probe strong-field gravity and test the limits of Einstein’s theory.

\begin{acknowledgments}
	The work is in part supported by NSFC Grant No.12205104, ``the Fundamental Research Funds for the Central Universities'' with Grant No.  2023ZYGXZR079, the Guangzhou Science and Technology Project with Grant No. 2023A04J0651 and the startup funding of South China University of Technology.
\end{acknowledgments}

\bibliographystyle{apsrev4-1}

\end{document}